\documentclass[conference, A4paper]{IEEEtran}
\IEEEoverridecommandlockouts

\usepackage{cite}
\usepackage{amsmath,amssymb,amsfonts}
\usepackage{algorithmic}
\usepackage{graphicx}
\usepackage{textcomp}
\usepackage{xcolor}
\usepackage{subfig}
\usepackage{url}
\usepackage{float}
\usepackage{multirow}
\def\BibTeX{{\rm B\kern-.05em{\sc i\kern-.025em b}\kern-.08em
    T\kern-.1667em\lower.7ex\hbox{E}\kern-.125emX}}
\usepackage{balance}

\newcommand{\etal}{\emph{et al.}}
\newcommand{\ie}{\emph{i.e.}, }
\newcommand{\eg}{\emph{e.g.}, }

\usepackage{booktabs}
\begin{document}

\title{Improving the Efficiency of VVC using\\Partitioning of Reference Frames

{}
\thanks{}
}

\author{
\IEEEauthorblockN{Kamran Qureshi\textsuperscript{1},
Hadi Amirpour\textsuperscript{1}, and Christian Timmerer\textsuperscript{1} }\vspace{0.2cm}

\IEEEauthorblockA{\textsuperscript{1} Christian Doppler Laboratory ATHENA, Alpen-Adria-Universitat, Klagenfurt, Austria}

}

\maketitle

\begin{abstract}
In response to the growing demand for high-quality videos, Versatile Video Coding (VVC) was released in 2020, building on the hybrid coding architecture of its predecessor, HEVC, achieving about 50\% bitrate reduction for the same visual quality. It introduces more flexible block partitioning, enhancing compression efficiency at the cost of increased encoding complexity.
To make efficient use of VVC in practical applications, optimization is essential. VVenC, an optimized open-source VVC encoder, introduces multiple presets to address the trade-off between compression efficiency and encoder complexity. Although an optimized set of encoding tools has been selected for each preset, the rate-distortion (RD) search space in the encoder presets still poses a challenge for efficient encoder implementations. In this paper, we propose Early Termination using Reference Frames (ETRF), which improves the trade-off between encoding efficiency and time complexity and positions itself as a new preset between medium and fast presets. The CTU partitioning map of the reference frames in lower temporal layers is employed to accelerate the encoding of frames in higher temporal layers. The results show a reduction in the encoding time of around 21\% compared to the medium preset. 
Specifically, for videos with high spatial and temporal complexities, which typically require longer encoding times, the proposed method achieves a better trade-off between bitrate savings and encoding time compared to the fast preset.
\end{abstract}

\begin{IEEEkeywords}
Versatile Video Coding, VVC, VVenC, Reference Frames, Partitioning
\end{IEEEkeywords}

\section{Introduction}
\label{intro} 

Video streaming services are constrained by network bandwidth limitations and server storage capacity, whilst users demand superior quality within existing network conditions. Standardization bodies have endeavored to meet the stringent expectations and demands of users. Versatile Video Coding (VVC)~\cite{bross2021overview}, the latest video coding standard finalized in 2020, achieves an average bitrate reduction of 50\% compared to its predecessor, HEVC~\cite{sullivan2012overview} while maintaining the same level of subjective quality. This improvement, however, comes at the cost of increased complexity, primarily due to the newly introduced block partitioning structure, known as  Quadtree with Multi-Type Tree (QTMT)~\cite{huang2021block}.

The current literature primarily aims to reduce the encoding complexity of VVC, focusing on both Intra- and Inter-frame optimizations. For Intra-frames, spatial features are utilized to optimize the rate-distortion (RD) search space~\cite{li2023texture, shang2023fast, yang2024fast, jiang2023low}. Some research studies aim to skip split modes such as BTH, BTV, TTH, and TTV~\cite{song2024fast}, while others use learning-based methods to predict the quadtree with nested multi-type tree (QTMT) depths to reduce encoding time~\cite{chen2023speed, kherchouche2024rd, li2024fastIntra}. Several studies aim to reduce the RD optimization burden for Inter-frames by leveraging temporal and spatial features. Kuang~\etal~\cite{kuang2022unified} use historical data from previously encoded CUs to predict the current CU's depth for both Intra- and Inter-frames. Shang~\etal~\cite{shang2023low} limit the number of modes (SKIP, Inter, affine, merge, Intra) evaluated for a CU block by considering the area and prediction modes of neighboring CUs. Li~\etal~\cite{li2024fast} employ a multi-decision-tree approach for partitioning decisions to optimize the RD search space for Inter-frames. Liu~\etal~\cite{liu2022light} propose a CNN model that divides a CTU into an $8\times8$ grid, using depth information within each grid for optimization. Peng~\etal~\cite{peng2023classification} develop a partition homogeneity map (PHM) to bypass RD cost searches for Inter-frames. Jiang~\etal~\cite{jiang2023extreme} utilize an extreme learning machine (ELM) algorithm for early split decisions. Lin~\etal~\cite{lin2024efficient} adopt a supervised contrastive learning algorithm to predict partitioning patterns. However, these studies evaluate their algorithms on the VVC Test Model (VTM), which is the non-optimized reference software for VVC.

For real-world applications, Fraunhofer HHI released an open-source \textit{optimized} software implementation of the VVC encoder, \ie VVenC~\cite{wieckowski2021vvenc}. Wieckowski~\etal~\cite{wieckowski2022vvc} integrated VVenC into Bitmovin's cloud-based solution, underscoring the importance of enhancing the efficiency of VVenC. VVenC has multiple presets: faster, fast, medium, slow, and slower, each with its own tradeoff between compression quality and encoding time. The slowest preset of VVenC achieves the same compression efficiency as VTM in less than half the time. The fastest preset of VVenC is $140\times$ faster than VTM, and the medium preset takes a third of the runtime compared to VTM~\cite{wieckowski2021vvenc}.

Although VVenC offers significant speed improvements over VTM, it retains the same complex block-based recursive partitioning structure used for RD cost optimization~\cite{ccetinkaya2021ctu, liu2022light}. Despite efforts to optimize encoding complexity, most existing research has focused on the VTM framework, targeting either Intra-frame~\cite{li2023texture, shang2023fast, yang2024fast, jiang2023low, song2024fast, chen2023speed, kherchouche2024rd, li2024fastIntra} or Inter-frame~\cite{kuang2022unified, li2024fast, liu2022light, peng2023classification, lin2024efficient} optimizations. This leaves the VVenC implementation relatively unexplored.

Recently, only a few studies have specifically focused on reducing encoding time for block partitioning in VVenC, including approaches like multi-rate encoding \cite{liu_preparing_2023}. While VVenC has made strides by incorporating software redesigns and SIMD optimizations to address performance bottlenecks, its overall complexity persists. The encoder's search algorithm and preset configurations have been fine-tuned based on \textit{Pareto optimal} configurations in the encoder’s design space~\cite{brandenburg2020towards}, yet further accelerating the encoder continues to be a challenging task.

In this paper, our focus is on enhancing the encoding efficiency of the widely used medium preset, which is already highly optimized. We propose a method named \textbf{E}arly \textbf{T}ermination using \textbf{R}eference \textbf{F}rames (\textbf{ETRF}), which aims to reduce the search space for videos, especially those with high complexity, by utilizing partitioning information from reference frames in lower temporal layers.

\section{Overview of VVC Partitioning}
\label{Overview}
VVC is based on the same block-based hybrid approach as its predecessors, where Intra/Inter prediction is followed by a 2D transform coding. VVC introduced novel tools that contribute to achieving approximately 50\% bitrate reduction compared to its predecessor, HEVC. Among other tools, the block partitioning structure~\cite{menon_incept_2021} provides significant coding gains~\cite{huang2021block}. 
In the VVC block partitioning structure, a frame of the video is divided into square-sized blocks of $128\times128$ pixels, known as the Coding Tree Units (CTUs). CTUs are recursively partitioned into Quad-Tree (QT) and Multi-Type Tree (MTT) Coding Units (CU). QT has square-shaped CUs, which can be further split into quad-split or MTT split until the minimum allowed depth. MTT has a rectangular shape and can only be further split into more MTTs until the minimum allowed depth is reached. The minimum allowed depth for a CU is $4\times4$. When a CU is split into QT, it contains four blocks, with each block having a quarter of the size. A CU may remain unsplit, which is referred to as a non-split (NS) partition. The MTT is made up of four shapes: Binary Tree Horizontal (BTH), Binary Tree Vertical (BTV), Ternary Tree Horizontal (TTH), and Ternary Tree Vertical (TTV)~\cite{huang2021block}.

In VVC, as opposed to HEVC, the concepts of CU, PU, and TU are simplified. In general, the size of CU, PU, and TU are the same, which significantly impacts the signaling overhead, except when subblock-based temporal motion vector prediction (SbTMVP)~\cite{yang2021subblock}, decoder-side motion vector refinement (DMVR)~\cite{gao2020decoder}, Intra subpartition (ISP)~\cite{de2019intra}, or subblock transform (SBT)~\cite{zhao2021transform} is performed. Under these circumstances, the CU or TU is divided into subblocks, and this side information might or might not need to be signaled.

\section{Early Termination using Reference Frames}
\label{ETRF}

The Random Access (RA) configuration effectively balances compression efficiency and random access, making it suitable for adaptive streaming and live broadcasting. In this configuration, frames in lower temporal layers are encoded first, serving as reference frames for higher layers. Although these reference frames are typically employed for motion compensation, the inherent similarity in the partitioning structure among collocated CTUs has not been fully exploited to optimize partitioning patterns. This method aims to leverage partitioning information from previously encoded frames in lower temporal layers to enhance the efficiency of encoding frames in higher temporal layers.

\begin{figure}[!t]
    \centering
    \includegraphics[width=0.99\linewidth]{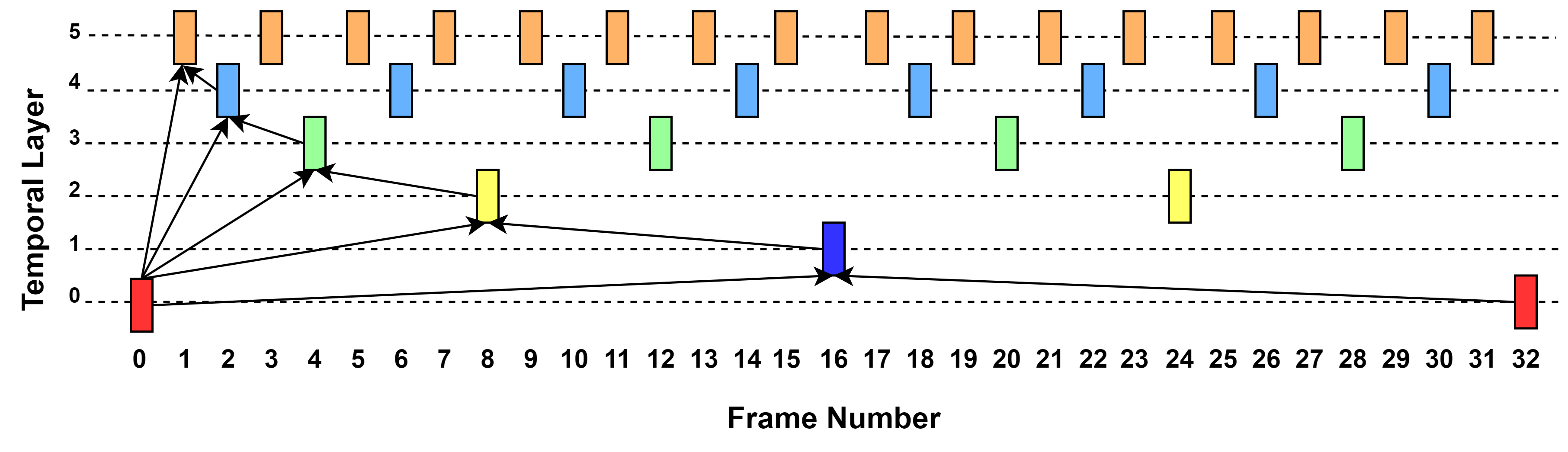}
    \caption{Temporal Layers for GOP 32.}
    \label{fig:temp-layers-gop32}
    \vspace{-1em}
\end{figure}

One of the main challenges is optimizing the search space due to the varying distance between the current frame and its nearest reference frames. In higher temporal layers, such as layers five and four (see Fig.~\ref{fig:temp-layers-gop32}), reference frames are one and two frames apart, respectively, resulting in high partitioning consistency. However, as temporal layers decrease, \eg layers 3, 2, 1, and 0, the distance increases to 4, 8, 16, and 32 frames, respectively. This greater distance reduces the correlation between collocated CUs, leading to partitioning errors that propagate, ultimately increasing bitrate.

To effectively control bitrate and improve encoding efficiency, the proposed ETRF method is applied to temporal layers two through five. By maintaining the distance between the current frame and its reference frames within eight frames, ETRF maximizes partitioning similarity and minimizes errors. This approach reduces the search space of a CU by referencing co-located CUs in the temporally closest frames. For frames in temporal layers two through five, RD cost calculations for partitioning above the defined bounds are skipped, and early termination is applied below the bounds to optimize CU processing.

\begin{figure}[!h]
\includegraphics[width=0.80\linewidth]{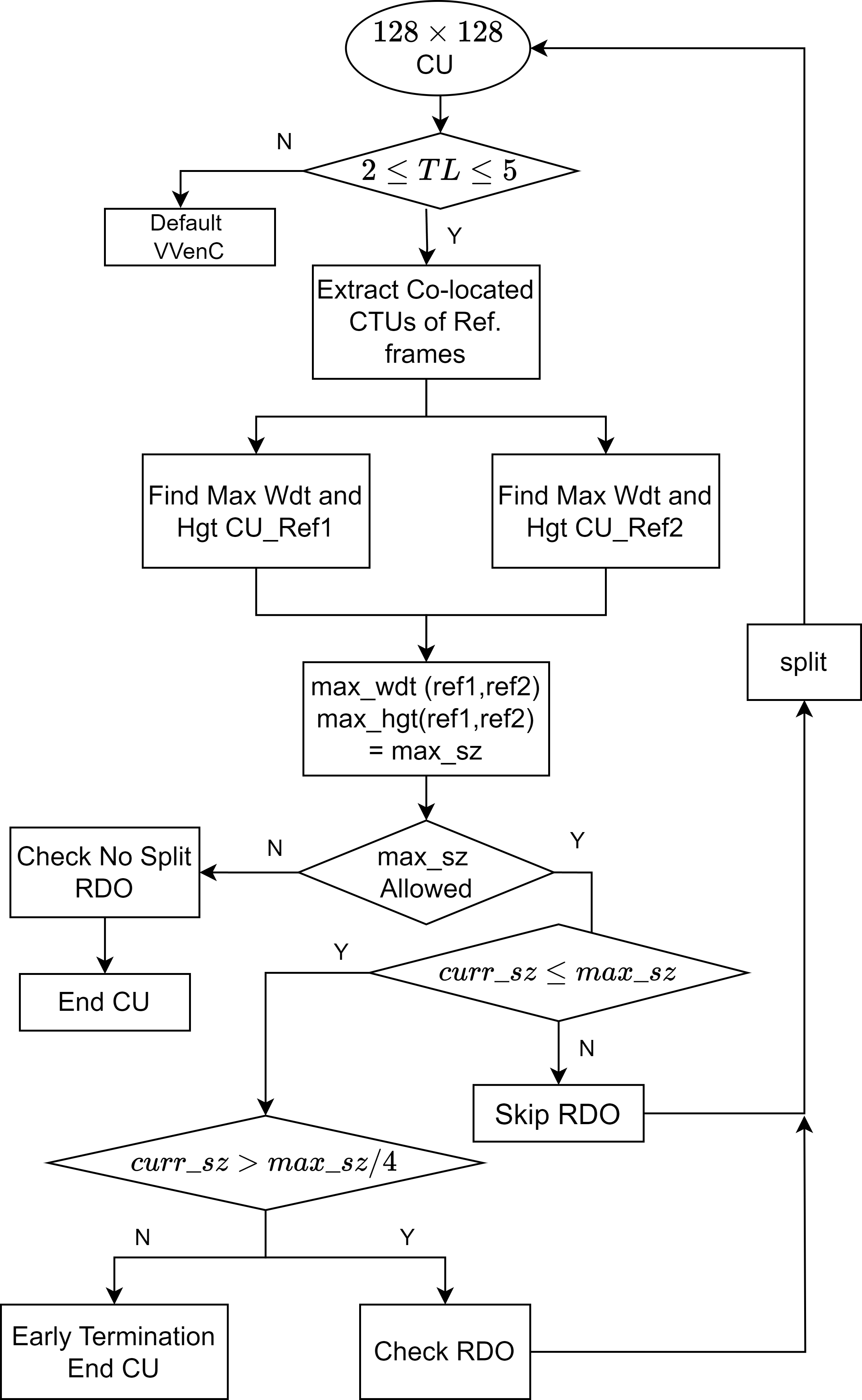}
\centering
\caption{Flowchart of the proposed ETRF method for VVC.}
\label{flow_diagram}
\vspace{-1em}
\end{figure}

The flowchart of ETRF is shown in Fig.~\ref{flow_diagram}. ETRF begins with recursively partitioning a CTU, starting with an initial size of $128\times128$. A check identifies whether the current frame belongs to the designated temporal layers. For frames outside these layers (\ie temporal layers 0 and 1), the default VVenC operation is applied. ETRF then extracts co-located CTUs from reference frames closest in time to the frame being encoded.

Partition information in a CTU includes parameters like width, height, starting coordinates, and QT and MTT depths, collectively referred to as features. ETRF defines a dynamic range for optimized partitioning by using the width and height of reference CUs (CU\_Ref1, CU\_Ref2). CU widths and heights in reference frames are stored in an $8\times8$ grid map, corresponding to the minimum QT depth in VVenC. Each CTU thus contains 256 values. Next, these 256 values are flattened and stored during encoding, with one row for width and another for height per CTU.

The search space for the current CU adapts based on the sizes of collocated CUs in reference frames, with the maximum width and height (max\_sz) calculated from these values. If max\_sz cannot be reached due to encoder optimizations preventing any split type, the fast partitioning method is bypassed, and the encoder defaults to its standard mode for that CU.

ETRF determines the minimum depth by allowing a depth margin instead of a hard boundary, as using a fixed boundary is impractical when max\_sz and min\_sz (minimum CU sizes) are identical for square CUs. The encoder permits recursive partitioning up to two depths below max\_sz, which is referred to as early termination. If max\_sz is less than or equal to $16\times16$, the early skip condition remains inactive, but higher depths are skipped due to the lower starting point. For max\_sz values greater than $16\times16$, the early skip condition bypasses all lower depths.
Early skipping is effective for lower depths when dependent frames are encoded at higher QP (lower quality), as these frames tend to match the CU depths of reference frames more closely.

\begin{figure}[!t]
\includegraphics[width=1\linewidth]{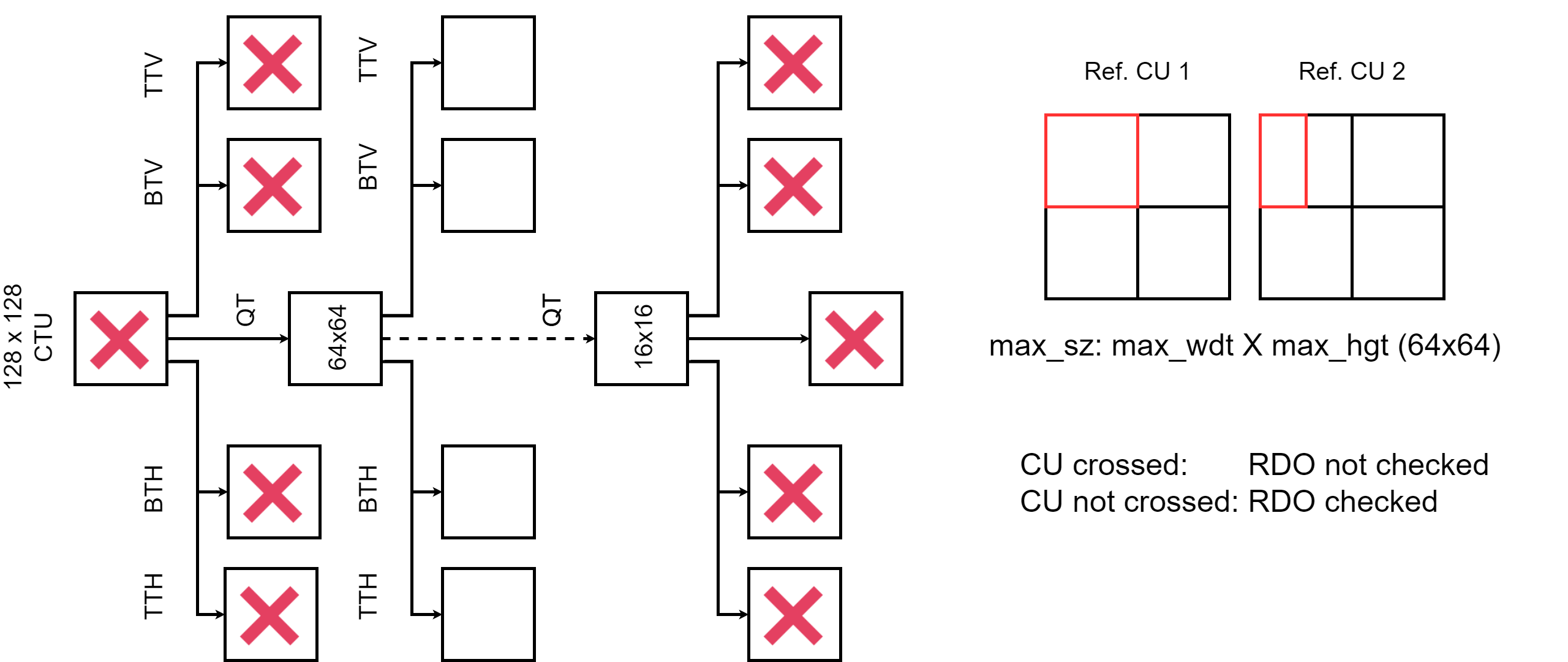}
\centering
\caption{An example of search space reduction for one CU. }
\label{search_space}
\vspace{-1em}
\end{figure}

Fig.~\ref{search_space} provides an example of our approach's skipping mechanism, reducing the search space for one CU and thus decreasing the complexity of RD optimization. In this example, reference frame CUs have a maximum width and height of $64\times64$. The dependent frame can search two depths lower to find the optimal partitioning pattern. This depth allowance accounts for the possibility of additional information in the same block due to object motion. Each CU undergoes this process, establishing its own search space boundaries.

\begin{table}[!h]
\centering
\caption{The performance of ETRF and the fast preset compared to the medium preset.}
\label{table_BDPSNR_BDBR}
\begin{tabular}{|c|c|c|c|c|}
\hline
\textbf{Class}                    & \textbf{Preset} & \textbf{BDT}   & \textbf{BDBR}  & \textbf{\rule{0pt}{10pt}$\frac{\textbf{BDBR}}{\textbf{BDT}}$\rule[-6pt]{0pt}{0pt}} \\ \specialrule{.12em}{.05em}{.05em}
\multirow{2}{*}{\textbf{A1}}      & ETRF            & 27.76          & 5.20           & 0.18              \\ \cline{2-5} 
                                  & Fast            & 61.16          & 41.11          & 0.67              \\ \hline
\multirow{2}{*}{\textbf{A2}}      & ETRF            & 25.28          & 6.79           & 0.26              \\ \cline{2-5} 
                                  & Fast            & 59.31          & 15.25          & 0.25              \\ \hline
\multirow{2}{*}{\textbf{B}}       & ETRF            & 23.30          & 4.82           & 0.20              \\ \cline{2-5} 
                                  & Fast            & 59.31          & 12.95          & 0.21              \\ \hline
\multirow{2}{*}{\textbf{C}}       & ETRF            & 22.90          & 2.35           & 0.10              \\ \cline{2-5} 
                                  & Fast            & 64.31          & 17.19          & 0.26              \\ \hline
\multirow{2}{*}{\textbf{D}}       & ETRF            & 15.24          & 1.33           & 0.08              \\ \cline{2-5} 
                                  & Fast            & 58.87          & 12.36          & 0.20              \\ \hline
\multirow{2}{*}{\textbf{E}}       & ETRF            & 12.09          & 3.28           & 0.27              \\ \cline{2-5} 
                                  & Fast            & 47.06          & 12.46          & 0.26              \\ \specialrule{.12em}{.05em}{.05em}
\multirow{2}{*}{\textbf{Average}} & \textbf{ETRF}   & \textbf{21.09} & \textbf{3.96}  & \textbf{0.18}     \\ \cline{2-5} 
                                  & \textbf{Fast}   & \textbf{58.33} & \textbf{18.55} & \textbf{0.31}  
                                  
                                  \\ \hline

\end{tabular}
\end{table}
\vspace{-1em}

\section{Experimental Results}
\label{exp.result}

\subsection{Setup}
\label{setup_l}

The ETRF method was evaluated using two datasets: the JVET Common Test Conditions (CTC) dataset~\cite{boyce_jvet-j1010_2018} and the Inter4K dataset~\cite{stergiou_adapool_2023}, which comprises a diverse set of videos with varying complexities. Each video was encoded on a server equipped with an Intel Xeon Gold 5220R CPU @ 2.2 GHz, utilizing a single thread for processing. Encoding tests were conducted across four quantization parameters (QPs): $22$, $27$, $32$, and $37$, ensuring a comprehensive evaluation of rate-distortion trade-offs.

The encoding experiments were performed using standalone VVenC version 1.7.0-rc~\cite{wieckowski_vvenc_2021}, comparing both the medium and fast presets under the random access configuration with a Group of Pictures (GOP) size of $32$. The proposed method builds on the medium preset, applying the same configuration while incorporating enhancements specific to ETRF.

To quantitatively assess the compression performance, we utilized the Bjøntegaard Delta Bit Rate (BDBR) metric, using Peak Signal-to-Noise Ratio (PSNR) as the quality measure. Additionally, time savings were evaluated through the Bjøntegaard Delta Time Saving (BDT) metric~\cite{herglotz_energy_2024}, which quantifies the reduction in encoding time while maintaining equivalent video quality. This dual-metric evaluation highlights both the bitrate efficiency and the computational savings of the proposed ETRF method.

Finally, $\frac{BDBR}{BDT}$ combines both compression efficiency and computational performance as an evaluation metric that offers a balanced perspective on the improvement in bitrate achieved per unit of time difference. Lower values indicate better overall performance.

\subsection{Results and Discussion}

In this section, we present a comprehensive performance evaluation of the proposed method, benchmarking it against two presets of VVenC. These presets correspond to configurations along the Pareto front, representing trade-offs between compression efficiency and computational complexity. Our comparison aims to highlight the improvements achieved by the proposed method in terms of both rate-distortion performance and encoding speed, providing a thorough analysis of its advantages over the existing VVenC presets.

\begin{figure}[tp]
    \centering
    \captionsetup{justification=centering}
    \subfloat[]{
        \includegraphics[width=0.50\linewidth]{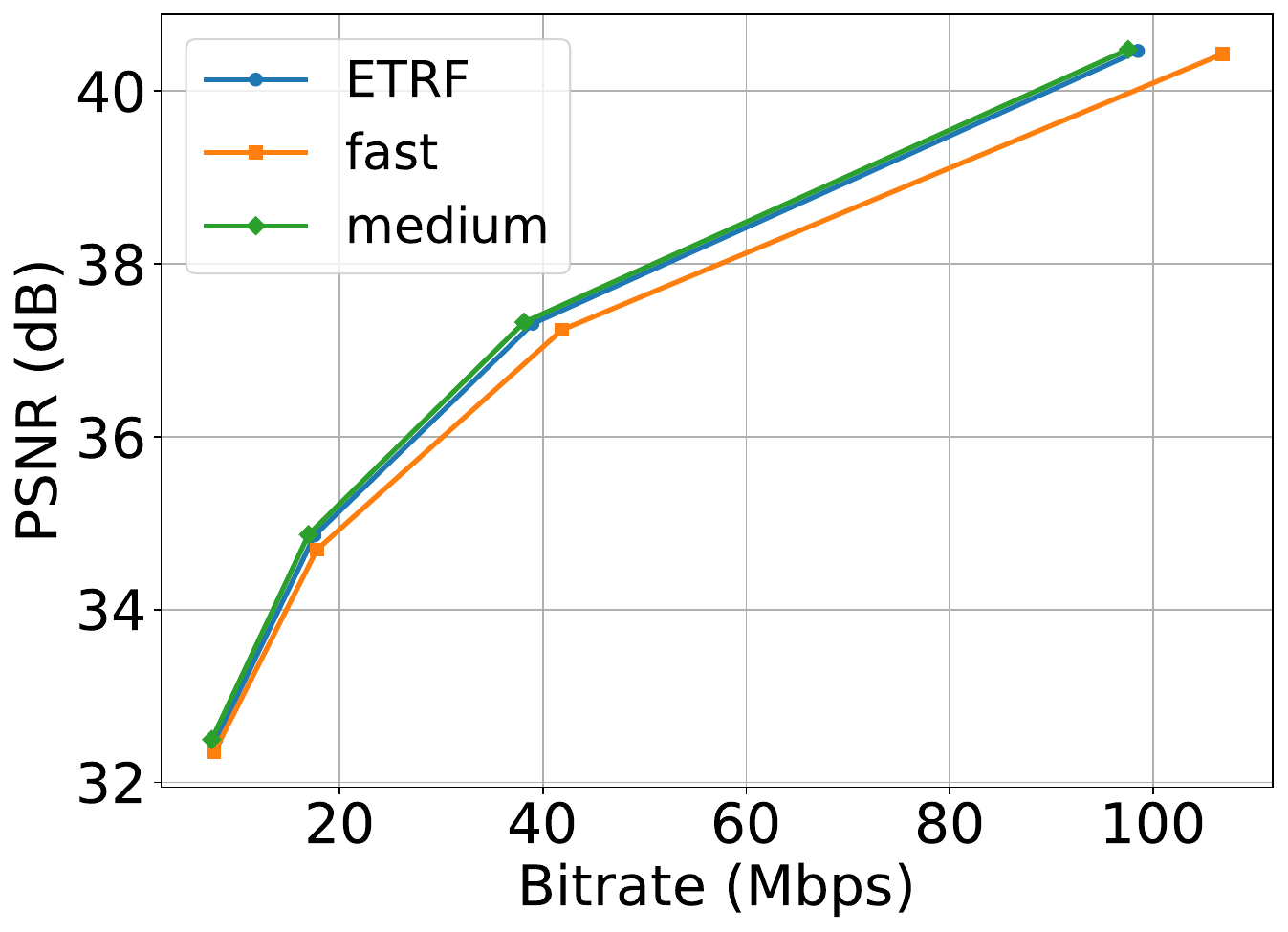}
        \label{bitrate_vs_psnr}
    }
    \subfloat[]{
        \includegraphics[width=0.50\linewidth]{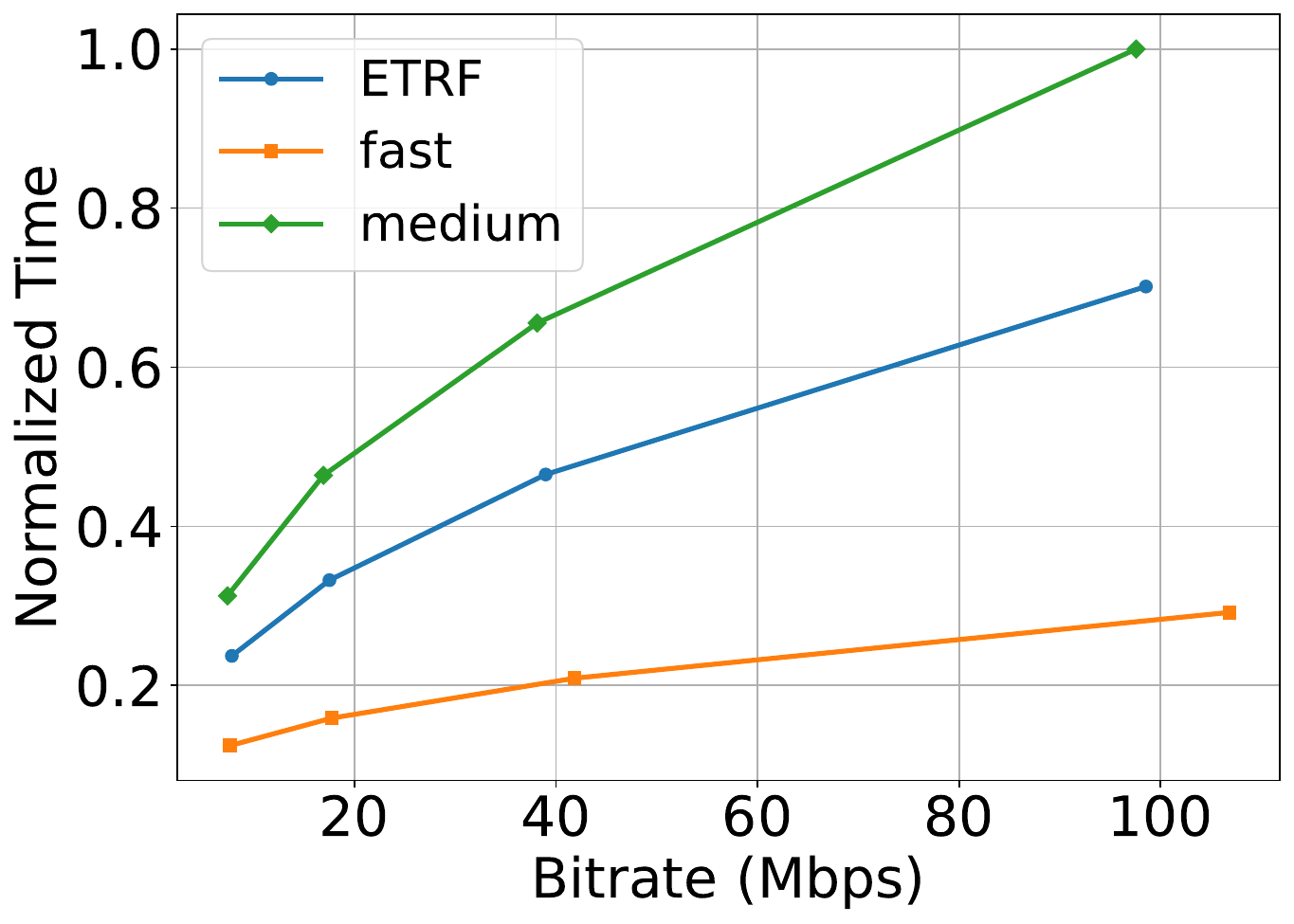}
        \label{bitrate_vs_time}
    }
    \caption{(a) PSNR vs. bitrate and (b) Time vs. bitrate of fast, medium, and ETRF preset.}
    \label{fig:subfigures_time_bitrate_psnr}
\end{figure}

Fig.~\ref{fig:subfigures_time_bitrate_psnr} shows the performance of medium, fast, and ETRF presets for an example video of the CTC dataset (video: ParkRunning3, 4k, 10bit, 50fps). As shown in Fig.~\ref{bitrate_vs_psnr}, the ETRF preset demonstrates an RD curve comparable to the medium preset while significantly outperforming the fast preset. However, as shown in Fig.~\ref{bitrate_vs_time}, it achieves this with a shorter encoding time.

To provide a more comprehensive comparison, considering both compression efficiency and encoding time, we calculate $\frac{BDBR}{BDT}$ of all classes for the CTC dataset. The performance of the ETRF preset is compared to the fast preset, using the medium preset as the reference, as shown in Table~\ref{table_BDPSNR_BDBR}. The results show that the ETRF preset achieves a lower $\frac{BDBR}{BDT}$ of 0.18, compared to 0.31 for the fast preset, indicating better overall efficiency. This means that ETRF offers a more balanced trade-off between quality and performance, making it a more efficient option for scenarios where both factors are critical.

\begin{figure}[tp]
    \centering
    \captionsetup{justification=centering}
    \subfloat[]{
        \includegraphics[width=0.50\linewidth]{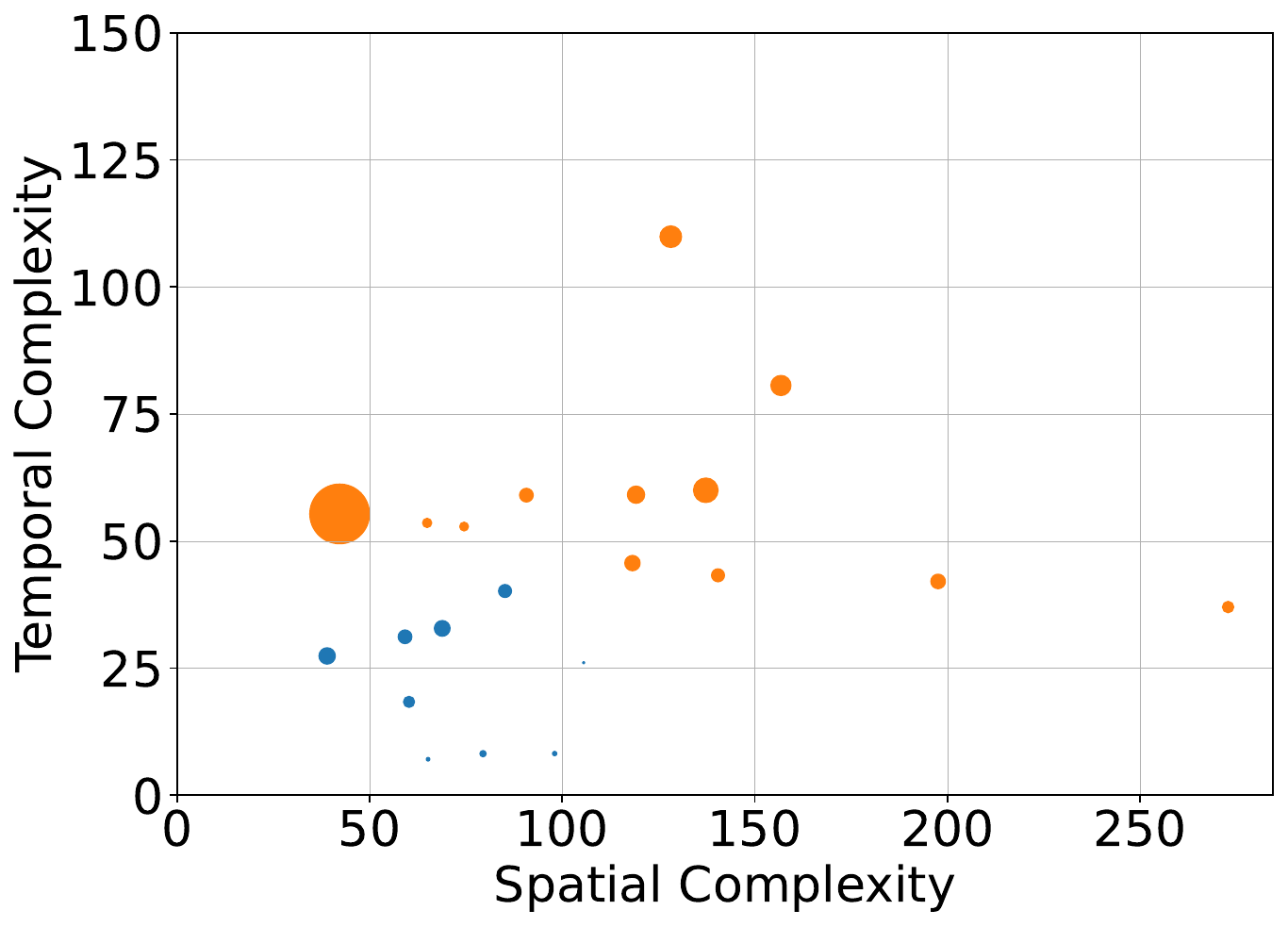}
        \label{ctc_spatial_temporal}
    }
    \subfloat[]{
        \includegraphics[width=0.50\linewidth]{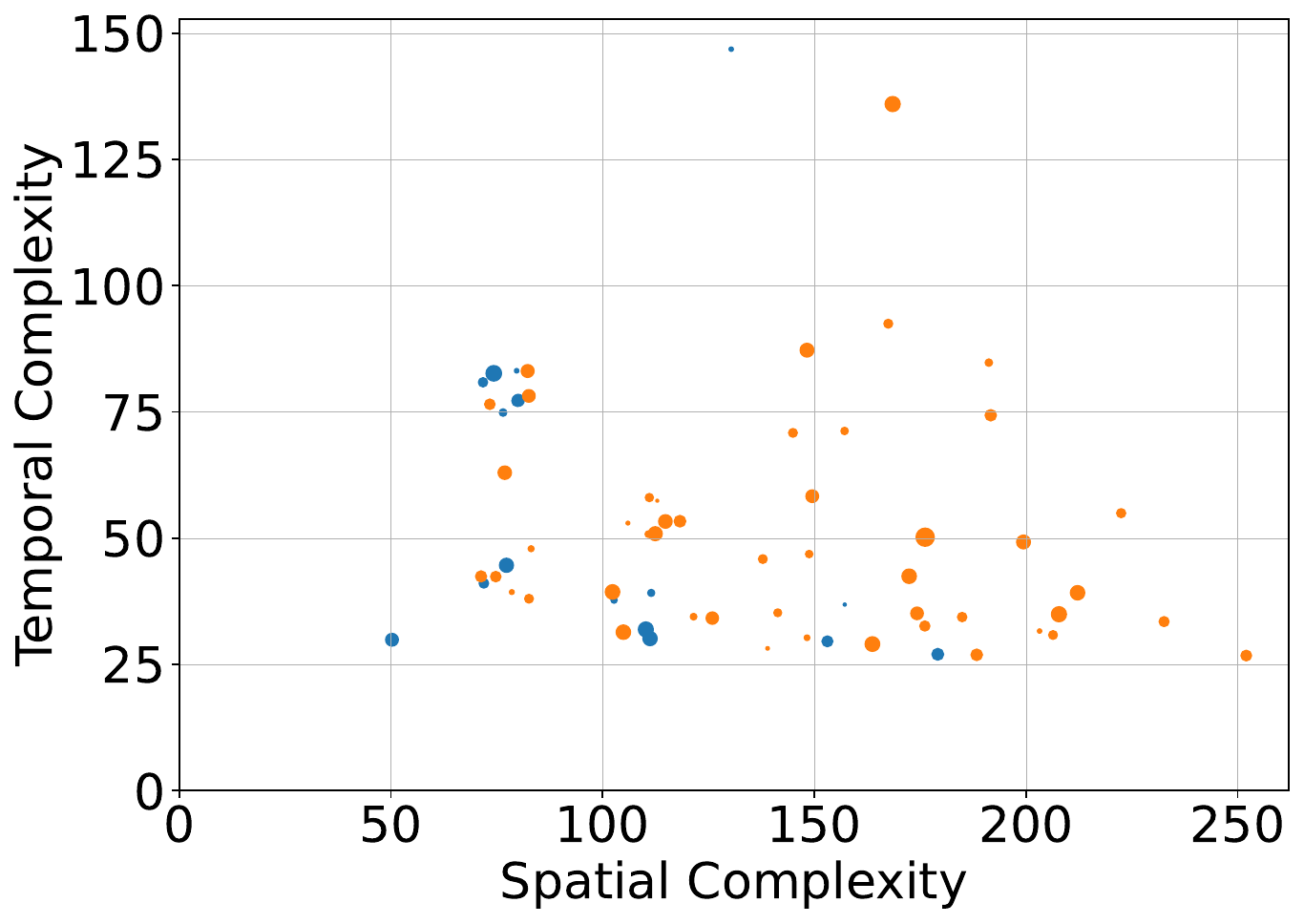}
        \label{inter4k_spatial_temporal}
    }
    \caption{$\frac{BDBR}{BDT}$ of (a) all ctc videos and (b) Inter4k videos relative to their spatial and temporal complexities.}
    \label{fig:subfigures_spatial_temporal}
\end{figure}

We further analyzed the results by examining $\frac{BDBR}{BDT}$ in relation to the spatial and temporal complexity of the videos. Fig.~\ref{ctc_spatial_temporal} presents the performance evaluation for all CTC videos, where $\frac{BDBR}{BDT}$ is used to assess performance based on spatial and temporal complexity. The complexity of each video, measured using VCA~\cite{menon2022vca}, is plotted along spatial (E) and temporal (h) axes. In the figure, orange dots represent videos with lower $\frac{BDBR}{BDT}$, indicating better performance, while blue dots correspond to videos with higher $\frac{BDBR}{BDT}$. The size of each dot reflects the magnitude of $\frac{BDBR}{BDT}$, either positive or negative. As shown in Fig.~\ref{ctc_spatial_temporal}, videos with higher spatial and temporal complexity generally exhibit better $\frac{BDBR}{BDT}$ performance compared to those with lower complexity. This is particularly advantageous, as high-complexity videos typically require more encoding time. Therefore, achieving greater time reduction in these cases is crucial for improving overall efficiency. 


Given the limited number of 4K videos in the CTC, we expanded our analysis to include videos from the Inter4k dataset. To ensure a fair comparison, 
we selected 50 videos from the similar spatial (E) and temporal (h) complexity ranges as the CTC videos \ie spatial range from approx [45:250] and temporal range from approx [25:150] as shown in Fig. \ref{inter4k_spatial_temporal}.
Most videos demonstrate an improved $\frac{BDBR}{BDT}$, represented by orange dots on the plot. A smaller subset of videos, marked by blue dots, show a decline in $\frac{BDBR}{BDT}$, highlighting specific trade-offs in certain video scenarios.
Comprehensive analysis using two datasets demonstrates that our method introduces a new preset (ETRF) with improved $\frac{BDBR}{BDT}$, achieving better compression efficiency for the same level of encoding time reduction.

\section{Conclusions}
\label{conc}
This paper introduces ETRF, a rapid CTU partitioning method that leverages partitioning patterns from temporally closest reference frames to optimize the rate-distortion (RD) search space.
ETRF enhances the RD search process by accounting for the similarity in partitioning of co-located CTUs in temporally closest reference frames present in the lower temporal layers.
The results show that ETRF is an effective intermediate preset between the medium and fast presets in VVenC. Notably, ETRF outperforms the fast preset in terms of $\frac{BDBR}{BDT}$ for the CTC dataset and high-complexity videos, which are particularly challenging to encode in real-time applications.

\section*{Acknowledgment}
The financial support of the Austrian Federal Ministry for Digital and Economic Affairs, the National Foundation for Research, Technology and Development, and the Christian Doppler Research Association, is gratefully acknowledged. Christian Doppler Laboratory ATHENA: \url{https://athena.itec.aau.at/}.

\balance
\bibliographystyle{ieeetr}
\bibliography{bibliography.bib,references_hadi}
\end{document}